\documentclass{ws-ijmpa}

\def\ra{\rightarrow}
\def\be{\begin{equation}}
\def\ee{\end{equation}}
\def\bea{\begin{eqnarray}}
\def\eea{\end{eqnarray}}
\def\g{\gamma}
\def\G{\Gamma}
\def\s{\sigma}

\newcommand{\e}{e^+e^-}
\newcommand{\EE}{e^++e^-}
\newcommand{\eb}{(e^+e^-)_b}
\newcommand{\m}{\mu^+\mu^-}
\newcommand{\MM}{\mu^++\mu^-}
\newcommand{\mb}{(\mu^+\mu^-)_b}

\newcommand{\ta}{\tau^+\tau^-}
\newcommand{\TT}{\tau^++\tau^-}
\newcommand{\tb}{(\tau^+\tau^-)_b}

\begin{document}

\markboth{A.A. Malik, I.S. Satsunkevich} {Production of $\tb$ in
electron positron collisions}

%
\catchline{}{}{}{}{}
%

\title{PRODUCTION OF $\tb$ IN ELECTRON POSITRON COLLISIONS}

\author{A.A. MALIK}

\address{Department of Physics, GC University,\\
Katchery Road, Lahore 54000, Pakistan\\
aliasimmalik@gcu.edu.pk}

\author{I.S. SATSUNKEVICH}

\address{B.I. Stepanov Institute of physics,
National Academy of Sciences of Belarus\\
F. Skoryna avenue 68, Minsk 220072, Belarus\\
satsunk@dragon.bas-net.by}

\maketitle

\begin{history}
\end{history}

\begin{abstract}
$\tb$ is an atom of simple hydrogenlike structure similar to
positronium $\eb$ and $\mb$. In this paper energy levels and decay
widths of different decay channels of $\tb$ are given. Cross section
of production of this atomic system in $\e$ annihilation taking into
account radiative corrections is calculated. According to our
estimates 886 $\tb$ atoms may be produced at BEPCII and 29 $\tb$
atoms are produced at VEPP-4M under the present experimental
conditions.

\end{abstract}

\ccode{PACS numbers: 36.10.-k, 14.60.Fg}

\section{Energy levels and decay channels of $\tb$}

$\tb$ is an electromagnetically bound state of $\tau^+$ and $\tau^-$
similar to positronium $\eb$ and $\mb$\cite{MSV}. It was
theoretically predicted\cite{Mof} in 1975 but never found
experimentally. QED is a well known theory to describe such type of
electromagnetically bound atomic states. As in the case of hydrogen
atom the hamilton operator for $\tb$ atom is \be H={{\bf P}^2\over
m_{\tau}}-{\alpha\over r}, \ee where $\alpha^{-1}(m_{\tau})\simeq
133.3$ is the constant of fine structure\cite{MS}, ${\bf
P}=-i\partial/\partial{\bf r}$, $r=|{\bf r}|$ is distance between
$\tau^+$ and $\tau^-$. This operator differs from the hydrogen atom
hamiltonian only by reduced mass. In hydrogen atom hamiltonian,
reduced mass is just the mass of electron but in case of $\tb$ atom
reduced mass is $m_{\tau}/2$, so the energy levels of $\tb$ are
obtained easily using the corresponding reduced mass. Different
corrections to the hamiltonian {\em eg.} relativistic mass growth,
orbital, spin-orbital, spin-spin and annihilation interactions, all
give the hyperfine structure levels. Perturbation hamiltonian
contains only the sum of spins of particles, so the energy levels
are divided into singlets (parastates with total spin 0) and
triplets (orthostates with total spin 1).

The energy levels of the $\tb$ atom are given as for hydrogen atom
with a reduced mass\cite{AMM} ${m_{\tau}/2}$ \be
E_n=-{m_{\tau}\alpha^2\over 4n^2}=-{25\over n^2}KeV. \ee Knowing the
energy spectra of this atom now we need to consider different decay
channels of $\tb$ system.

There are two classes of decay channels\cite{Per}. In the first
class the $\tau^-$ or $\tau^+$ decay through the weak interaction in
the normal way and the atomic state disappears. The decay width is
\be \G(\tb,\tau\ decay)={2\over\tau}= 4.53\times 10^{-3}eV, \ee
where $\tau$ is lifetime of tau lepton.

In the second class of decay channels the $\tau^-$ and $\tau^+$
annihilate. The annihilation requires that the atomic wave function
$\Psi (r)$ be unequal to 0 at r=0 {\em i.e.} $\Psi(0)\neq0$. Here r
is the distance between the $\tau^-$ and $\tau^+$. Therefore in
lowest order, annihilation occurs only in L=0 state, {\em i.e.} S
state. The annihilation channels of $n^{3}S_{1}$ state of $\tb$ with
corresponding decay widths are following: The channel \be
\tb\ra\g+\g+\g, \ee has the width \be
\G(\tb\ra3\g)={2(\pi^2-9)\alpha^6m_{\tau}\over9\pi
n^3}=1.95\times10^{-5}eV. \ee The two channels \be \tb\ra\EE, \ee
\be \tb\ra\MM, \ee have the same width \be
\G(\tb\ra\EE)={\alpha^5m_{\tau}\over6n^3}=7.04\times10^{-3}eV. \ee
Finally there is hadron channel \be \tb\ra hadrons. \ee

For $\tb$ we can calculate the width of this channel using colliding
beams $\e$ annihilation data at $E_{tot}\sim2m_{\tau}$ \be \s(\EE\ra
hadrons)\approx2\s(\EE\ra\MM). \ee Therefore \be \G(\tb\ra
hadrons)\approx2\G_{ee}=1.41\times10^{-2}eV. \ee Neglecting
$\G(\tb\ra3\g)$ we get the total width \be \G_n\approx\G(\tb,\tau\
decay)+4\G_{ee}=\Biggl(4.5+{28.1\over n^{3}}\Biggr)\times10^{-3}eV.
\ee Using these values we can calculate production cross section of
$\tb$.

\section{Production cross section of $\tb$}

$\tb$ can be produced in $\e$ annihilation just below $\tau$ pair
threshold \be \EE\ra\g_{virtual}\ra\tb \ee Corresponding production
cross section according to the Breit-Wigner equation, is \be
\s(\EE\ra\tb)={3\pi\over4m_{\tau}^2}{\G_{ee}\G\over(E-2m_{\tau})^2+\G^2/4}.
\ee Here E is the total energy of $e^{-}$ and $e^{+}$. The peak
cross section is \be \s(\EE\ra\tb,peak)={3\pi\over
m_{\tau}^2}{\G_{ee}\over\G}=0.5mb.\ee Taking into account the
radiative corrections\cite{Bay} we get \be
\s_r=\s\exp\Biggl(-{4\alpha\over\pi}ln{m_{\tau}\over
m_e}ln{m_{\tau}\over\G}\Biggr). \ee

Now we consider the possibility that initial $e^{-}$ or $e^{+}$ can
radiate soft photon with \be {\omega\over m_{\tau}}\ll 1. \ee The
criteria of perturbative theory application in this case are \be
\alpha ln{m_{\tau}\over m_e}\ll 1,\hspace{1cm}
{\alpha\over\pi}ln{m_{\tau}\over
m_e}ln{m_{\tau}\over\omega_{min}}\ll 1.\label{eq:cond} \ee From
(\ref{eq:cond}) it implies that we can put $\omega_{min}$ equal to
the decay width of $1^{3}S_{1}$ state of $\tb$. The cross section of
radiative production of $\tb$ is \be
\s_{rad}=\sum_n{\pi\over2n^3}{\alpha^6\over
m_{\tau}\omega_{min}}\Biggl(2ln{m_{\tau}\over m_e}-1\Biggr). \ee

As the electrons and positrons in accelerator beams have some energy
spread so we should use an average value of cross section\cite{Cab}
\be <\s>=\int\s(E)\rho(E)dE. \ee

\section{Production of $\tb$ at BEPCII}

BEPCII is an electron positron Collider working in the tau-charm
energy region at IHEP, Beijing, China. It has only one detector
BESIII\cite{Wan}. It is an excellent machine to produce $\tb$. Let
the electrons and positrons are equally distributed in some energy
interval $\triangle E$ near threshold. At BEPCII we have $\triangle
E=1.4MeV$\cite{Mo} \be <\s_r>=\s_r{\pi\over2}{\G_{\tau
ee}\over\triangle E}=0.70pb, \ee \be
<\s_{rad}>=\s_{rad}{\omega_{min}\over\triangle E}\Biggl(ln{\triangle
E\over\omega_{min}}-1\Biggr)=13.36pb. \ee Number of events which
will be produced at BEPCII having luminosity $L=63pb^{-1}$ are \be
N=(<\s_r>+<\s_{rad}>)L=14.06\times63=886events. \ee If suitable
event selection criteria adopted $\tb$ can be observed using BESIII.
The cross section $\s(\e\ra\ta)$ just below $\tau$ pair threshold at
3.5538 GeV, with the account of all the radiative corrections
is\cite{Kir} \be \s(\EE\ra\TT)=110pb. \ee Branching ratio for
$\tau\ra\mu\nu\nu$ is 0.1735, so we have \be
\s(\EE\ra\TT\ra\MM)=110\times(0.1735)^2=3.3pb. \label{free} \ee
Branching ratio for $\tb\ra\m$ is 0.21, so the cross section
is\cite{MSQS} \be \s(\EE\ra\tb\ra\MM)=14.06\times0.21=3pb.
\label{bound} \ee Such a big cross section is due to the branching
ratio for $\tb\ra\MM$ which is 0.21 in spite of 0.03 as for free
$\ta$ pair. Comparing ``(\ref{free})'' and ``(\ref{bound})'' we see
that $52\%$ of $\mu\mu$ events are from free $\ta$ pair and $48\%$
$\mu\mu$ events from $\tb$. Same will be true for ee events although
there will be no contribution of $\tb$ to $e\mu$ events.

\section{Production of $\tb$ at VEPP-4M}

VEPP-4M is another electron positron Collider working in the 1-6 GeV
energy region at Budker Institute of Nuclear Physics, Novosibirsk,
Russia. A precise $\tau$ lepton mass measurement performed at the
VEPP-4M collider with the KEDR detector has been reported
recently\cite{Ana}. As in previous experiments the contribution of
the bound state $\tb$ has not been taken into account. Let us
calculate how much $\tb$ were produced in this experiment. We asume
that the electrons and positrons are equally distributed in some
energy interval $\triangle E$ near threshold at VEPP-4M we have
$\triangle E=1.07MeV$. Then $<\s_r>=0.92pb$ and
$<\s_{rad}>=17.20pb$. So the number of $\tb$ produced at VEPP-4M
having integrated luminosity $L=1.605pb^{-1}$ at $<E>=1776.896MeV$
is \be N=(<\s_r>+<\s_{rad}>)L=18.12\times1.605\simeq29events. \ee
These $\tb$ will contribute to ee and $\mu\mu$ events. It is
reported\cite{Ana} that at $<E>=1776.896MeV$ 6 $\tau\tau$ events
were detected, this number will ultimately increase if we take into
account $\tb$ and it will be a direct evidence for the existence of
$\tb$.

\section{Conclusion}

It follows from our estimates for $\tb$ decay rates and production
cross section that $\tb$ decays give a rather big number for ee and
$\mu\mu$ events at BEPCII and at VEPP-4M.

\section*{Acknowledgements}

Work of A.A. Malik was supported by Higher Education Commission of
Pakistan under the grant No. 1-5/90/HEC/R\&D/2006.



\begin{thebibliography}{00}  

\bibitem{MSV} A.A. Malik and I.S. Satsunkevich, {\it Vesti NANB Ser. fiz.-mat. nav.} No 3, 78 (2001).
\bibitem{Mof} J.W. Moffat, {\it Phys. Rev. Lett.} {\bf 35}, 1605 (1975).
\bibitem{MS} W.J. Marciano and A. Sirlin {\it Phys. Rev. Lett.} {\bf 61}, 1851 (1988).
\bibitem{AMM} C. Avilez, R. Montemayor and M. Moreno, {\it Lettere Al Nuovo Cimento Series 2} {\bf 21}, 301 (1978).
\bibitem{Per} M.P. Perl, SLAC-PUB-6025 (1992).
\bibitem{Bay} V.N. Bayer and V.S. Sinakh, {\it ZhETF} {\bf 41}, 1576 (1961).
\bibitem{Cab} N. Cabibbo and R. Gatto, {\it Phys. Rev.} {\bf 124}, 1577 (1961).
\bibitem{Wan} Y.F. Wang, {\it Int. Jour. Mod. Phys. A} {\bf 21}, 5371,
(2006).
\bibitem{Mo} X.H. Mo, {\it Nucl. Phys. Proc. Suppl.} {\bf 169}, 132,
(2007).
\bibitem{Kir} J. Kirkby, CERN-PPE/96-112.
\bibitem{MSQS} A.A. Malik and I.S. Satsunkevich in {\it Proc. of 3rd Int. Workshop on Quantum Systems}, (Minsk June 9-13 1999), p. 104.
\bibitem{Ana} V.V. Anashin {\it et al.}, {\it PZhETF} {\bf 85}, 429
(2007).

\end{thebibliography}
\end{document}